\documentclass[preprint2]{aastex}
\usepackage{amsmath,amsfonts,amssymb,graphicx,graphics,color,times,bbm,psfrag,dsfont,slashed}
\usepackage[latin9]{inputenc}
\usepackage{pifont}
\usepackage{natbib}
\usepackage{subfigure,epsfig,epstopdf}
\usepackage{bm}
\usepackage{framed}
\usepackage{enumerate}
\newcommand{\be}{\begin{equation}}
\newcommand{\ee}{\end{equation}}
\newcommand{\bea}{\begin{eqnarray}}
\newcommand{\eea}{\end{eqnarray}}

\begin{document}
%\preprint{}

\title{Torsional oscillations of nonbare  strange stars}

\author{Massimo Mannarelli and Giulia Pagliaroli}

\affil{INFN, Laboratori Nazionali del Gran Sasso, Via G. Acitelli, 22, I-67100 Assergi (AQ), Italy}
\email{massimo@lngs.infn.it}

\author{Alessandro Parisi, Luigi Pilo and Francesco Tonelli}
\affil{Dipartimento di Scienze Fisiche e Chimiche, Universit\`a di L'Aquila, I-67010 L'Aquila, Italy and INFN, Laboratori Nazionali del Gran Sasso, Via G. Acitelli, 22, I-67100 Assergi (AQ), Italy}

\begin{abstract}
Strange stars are one of the possible compact stellar objects that can be formed after a supernova collapse.  
We consider a model of strange star having an inner core in the  color-flavor locked phase surmounted by a crystalline color superconducting layer.  These two phases constitute the {\it quarksphere}, which we assume to be the largest and heaviest part of the strange star.  The next layer consists of standard nuclear matter forming a ionic crust, hovering on the top of the quarksphere and prevented from falling  by a strong dipolar electric field. The dipolar electric field arises because quark matter is confined in the quarksphere by the strong interaction, but electrons can leak outside forming a few hundreds Fermi thick electron layer separating the ionic crust from the underlying quark matter.  The ionic matter and the crystalline color superconducting matter constitute two electromagnetically coupled crust layers.  We study the torsional oscillations of these two layers. Remarkably, we find that if a fraction larger than $10^{-4}$  of the energy of a Vela-like glitch is conveyed to a torsional  oscillation, the ionic crust will  likely break. 
The reason is that the very rigid and heavy crystalline color superconducting crust layer will  absorb only a small fraction of the glitch energy, leading  to a large amplitude torsional oscillation of  the ionic crust.
\end{abstract}


\section{Introduction}
The properties of hadronic matter at densities larger than the nuclear saturation  density are mostly unknown. Even though much effort has been devoted to the study of very dense hadronic matter, we still do not know the actual ground state of matter as a function of the baryonic density. In particular, we are still unable to prove or falsify the hypothesis advanced by Bodmer and Witten \citep{Bodmer:1971we,Witten:1984rs} that standard nuclei are not the ground state of matter. According to this hypothesis  the energetically favored ground state of baryonic matter could be  a {\it collapsed  state}:  an hadronic configuration corresponding to a (not yet experimentally observed) short-range free-energy minimum of the strong interaction. For small baryonic numbers, the collapsed state  can be thought as a droplet of  quarks and gluons  having the size of few Fermi, smaller than the ten Fermi size characteristic of standard nuclei.  The droplet has a surface tension given by the bag pressure,  which keeps the matter density inside the droplet at almost a constant value  larger than the saturation density of standard nuclear matter. Therefore, with increasing  baryonic number, $A$, the size of the droplet grows with $r\sim A^{1/3}$, as characteristic of self-bound objects. In the following we will  assume that for any baryonic density the quark droplet corresponds to the free-energy minimum of the system, meaning that it is not possible to minimize the free energy by fission processes. By contrast, standard nuclear matter can only form small clumps corresponding to fission stable nuclei. 

At vanishing temperature and considering that the only effect of the strong interaction is a bag pressure, the collapsed state can be thought as consisting of almost free quarks filling the  pertinent energy levels up to the corresponding Fermi energy. When the light quark chemical potential exceeds the strange quark mass, the strange quark states start to be  populated by means of weak decay processes. In this case, the  collapsed state corresponds to ``catalyzed" $u, d, s$ quark matter~\citep{Witten:1984rs}, the so-called strange matter.  A very big clump of  strange matter is called a strange star~\citep{Alcock:1986hz, Haensel:1986qb}. Strange stars and neutron stars can be viewed as two different end products of supernova explosions and are classified as  compact stellar objects (CSOs), which are observed as stars having the radius of about $10$ km and the mass of about a solar mass, $M_\odot$. Actually,  assuming that strange matter is  self-bound implies that  the  size of a strange star has no lower bound, meaning that  strange stars much smaller than standard neutron stars  can exist. Gravity plays a role only for very massive objects, restricting the  mass to about $2 M_\odot$ (for a sufficiently stiff equation of state (EoS)), see for example~\cite{Mannarelli:2014ija}.

Assuming that the ground state of hadronic matter consists of  strange matter, it still does not clarify unambiguously the properties of the system.  The largest  value of the quark chemical potential that can be reached in massive strange stars is of the order of $400-500$ MeV. Even considering this extreme case,  the strong interaction is still nonperturbative and quantum chromodynamics (QCD)  is not under quantitative control. Therefore, approximation schemes must be used. Analyses using various models indicate that at the densities relevant for strange stars,  deconfined and cold quark matter is likely in a  color superconducting (CSC) phase, see~\cite{Rajagopal:2000wf, Alford:2007xm, Anglani:2013gfu} for reviews,  in which quarks form Cooper pairs breaking  the $SU(3)_\text{color}$ gauge symmetry. The reason is that the  estimated critical temperature of color superconductors is at least of the order of few MeV,  much larger than the  tens of KeV temperature of  few seconds old CSOs. Thus, if quark matter is present in CSOs, it should be in a  CSC phase.

The CSC phase is actually a collection of phases and pinning down the favored quark pairing  is not trivial.
Using  models based on one gluon exchanges or on  instanton exchanges, it can be shown that at asymptotic densities the  color-flavor locked (CFL) phase~\citep{Alford:1998mk} is energetically favored. In the CFL phase $u, d, s$ quarks of all colors pair coherently maximizing the free-energy gain. Although this phase  is very robust, it might be  that strange stars  do not completely consist of CFL matter. The reason is that when  the effective strange quark mass has a value  comparable with the quark chemical potential, a considerable free-energy penalty results for producing strange quarks. If the free-energy penalty is larger than the free-energy gain associated to the CFL pairing ---exceeding the corresponding Chandrasekhar-Clogston limit \citep{Chandrasekhar,Clogston,Anglani:2013gfu}--- a different phase is favored.  In the present paper we assume that the CFL phase is realized in the central  and denser part of the strange star and the next favored phase  down in density is the crystalline color superconducting (CCSC) phase~\citep{Alford:2000ze,Anglani:2013gfu}, which is then  realized in the outer and less dense part of the strange star~\citep{Rupak:2012wk,Mannarelli:2014ija}. The CCSC phase is characterized by a periodic modulation of the diquark pairing. For our purposes, the most important property of the CCSC phase is that this periodic modulation is mechanically rigid, with an extremely large shear modulus~\citep{Mannarelli:2007bs,Anglani:2013gfu}. In our  model the CFL phase and the CCSC phase  form the {\it quarksphere}  comprising most of the mass of the strange star. It is known by various model calculations, see for example~\cite{Anglani:2013gfu}, that the CCSC phase is the favored phase for sufficiently mismatched quark Fermi spheres, corresponding to a quark chemical potential in a certain range of values. The actual matter density at which the CCSC is favored depends on the detailed dependence of the strange quark mass on the quark chemical potential and the corresponding value of the CFL gap parameter. Since these quantities cannot be precisely computed, we will assume that at a certain radius, $R_\text{CFL}$, there exists a boundary between the CFL and the CCSC phase.  

An important aspect of  dense quark matter in CSOs is that it must be electrically neutral. The typical combined effect of charge neutrality, $\beta$ decays and nonvanishing strange quark mass, $M_s$, is to populate the electron states, see for example~\cite{Alford:2001zr}. Qualitatively, the reason is that a large strange quark mass disfavors the appearance of strange quarks by light quark $\beta$ decay, thus the electrical neutrality condition is satisfied by populating electron states.    A notable exception  is the CFL phase, in which the symmetric pairing induced by the strong interaction forces equal number of $u, d$ and $s$ quarks, thus no electrons are present. On the other hand, in the CCSC phase, the quark Fermi momenta are mismatched and electrons are needed to maintain the electric charge neutrality. Therefore, our model of strange star consists of  a central electron-free region of CFL matter in contact with an electron-rich region of CCSC matter.  The presence of electrons in the CCSC phase leads to an interesting phenomenon happening  at the surface of the strange star where the strong interaction confines quarks beneath the CCSC strange star surface. Electrons are  only bound by the electromagnetic force and will  typically spread outside the quarksphere for a length-scale of the order of the Debye screening length,  of hundreds of Fermi. A CSO made by quark matter surrounded by an electrosphere is called a bare strange star, see~\cite{Alcock:1986hz} for more details.

In some circumstances it is possible that a standard ionic crust hovers on the top of the strange star. The reason is that in the formation process, or by accretion, the strange star attracts hadronic matter.  Neutrons are absorbed by the quarksphere, but  ions are repelled by the positively charged quark surface. Thus, ions hover on the top of the strange star as far as the repulsive electric force and the gravitational force balance. If the accreted material is enough, it will  form a ionic crust separated by the  deconfined quark matter surface by a thin  electron layer. This CSO is called a nonbare strange star~\citep{Alcock:1986hz}. In the following we will  consider a model of nonbare strange star having two crusts, an inner crust layer consisting of CCSC matter and an outer crust layer consisting of standard ionic matter. 

In~\cite{Mannarelli:2014ija} we studied the torsional oscillations of bare strange stars. In the present paper we extend that study to nonbare strange stars, analyzing the torsional oscillations of the two coupled crust layers.  An intriguing aspect of this model is that these two crusts have very different shear moduli and densities, therefore nontrivial dynamical process can happen. We focus our analysis on $\ell=1$ modes, corresponding to oscillatory twists of the crust. These modes  do not conserve angular momentum, therefore we assume that they are triggered by events that  transfer angular momentum to the strange star crust. A typical event of this sort is a stellar glitch. For definiteness,  we will assume that an energy of the order of the one released in a Vela-like glitch is conveyed to  the $\ell=1$  modes; we will show how our results can  be appropriately rescaled if a different energy scale is used.  One of the most interesting results that we obtain is that the shear strain has a radial dependence with a maximum  much closer to the star surface than in standard neutron stars.  Moreover, in standard neutron stars the energy of the torsional oscillations is spread across the entire  crust, which is  more than a km thick. In the considered model of nonbare strange stars, for a sufficiently thin CCSC crust,   all the energy of the torsional oscillation is conveyed in a layer few hundred meters thick,  corresponding to the ionic crust at densities below neutron drip. Therefore, the ionic crust layer will likely crack even if  a small fraction, of order  $10^{-4}$, of the Vela-like glitch is conveyed to the ionic crust.

The present paper is organized as follows. In Sec.~\ref{sec:general} we present  a general description of the nonbare strange star  model. In Sec.~\ref{sec:background} we determine the stellar structure by solving the hydrostatic equilibrium equations. In Sec.~\ref{sec:torsional} we discuss the torsional oscillations of the stellar crust. In Sec.~\ref{sec:conclusions} we draw our conclusions. In Appendix~\ref{sec:rigid_slab} we discuss a simple toy model of two rigid slabs, determining the corresponding torsional eigenfrequencies.

\section{General description}\label{sec:general}

\begin{figure}[t!]
\includegraphics[width=8.cm]{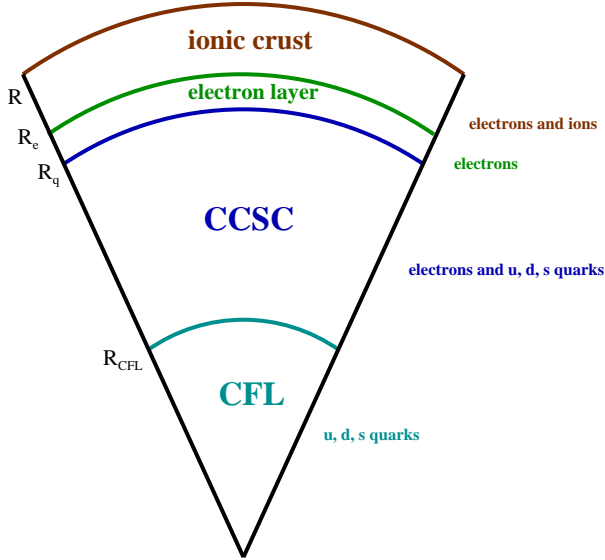}
\caption{Pictorial description of the strange star structure (not in scale). The stellar radius of the considered model is $R\simeq 9.18$ km.   Most of the strange star, for $0<R<R_\text{q} \simeq 8.98$ km, consists of $u, d, s$ quark matter in a color superconducting phase, forming the quarksphere.  The inner  part of the quarksphere, for  $0<R<R_\text{CFL}$,  consists of electron-free    color-flavor locked  $u, d, s$  matter;  the outer part of the quarksphere, $R_\text{CFL}<R<R_\text{q}$,   consists of the electron-rich    crystalline color superconducting  $u, d, s$  matter.  The radius $R_\text{CFL}$ is unknown and we consider values in the range $0\leq R_\text{CFL} \leq R_\text{q}$. On the top of the quarksphere the  electron layer extends between $R_\text{q}$  and $R_\text{e}$; it is  just few hundreds of Fermi thick and consists of the electrons leaked outside the quarksphere. The last layer is the  ionic crust, which has the same properties of standard neutron star crust.  \label{fig:star}}
\end{figure}

The considered nonbare strange star model is depicted in Fig.~\ref{fig:star}.  The   $u, d$ and $s$ quark matter is radially confined by the strong interaction in the quarksphere, with radius $R_q$. This part of the  star can be considered as a big hadron with an extremely large baryonic number. Since the quark chemical potential is large and the temperature is sufficiently low, deconfined quark matter is assumed to be in a CSC phase. The inner part  of the quarksphere, up to the radius $R_\text{CFL}=a R_q$, with $a \leq 1$, is in the  CFL phase.  The CFL phase is surmounted by the  CCSC phase, extending between $R_\text{CFL}$ and $R_q$. The outer part of the nonbare strange star consists of a thin electron layer, extending between $R_q$ and $R_e$. This region is only a few hundreds of Fermi thick~\citep{Alcock:1986hz}, thus when considering the  hydrostatic equilibrium configuration we will  not distinguish between  $R_q$ and $R_e$. 

The next layer on the top of the electrosphere is the ionic crust, which is composed by ions and electrons. This layer extends between $R_e$ and $R$, reaching at most the neutron drip density~\citep{Alcock:1986hz}. Densities above the neutron drip point cannot be attained because the ionic crust is maintained from collapsing on the underlying strange star surface by the electrostatic surface field. If the density of the crust reaches the neutron drip point, neutrons are liberated falling on the underlying strange star surface by gravitational attraction and are   thus absorbed in the quarksphere. This process puts a  limit on the maximum mass of the ionic crust~\citep{Alcock:1986hz}. The ionic crust layer of nonbare strange stars has the same mechanical properties of the outer crust of standard neutron stars, meaning that we will  considered the known EoS of nuclear matter below neutron drip and the estimated values of the nuclear matter elastic moduli. Moreover, as we will discuss below, the ionic crust is in part liquid forming the so-called ocean.

One important aspect to clarify is that for the  background  configuration we assume hydrostatic equilibrium, but oscillations are described considering an elastic response to the mechanical stress. Although the descriptions of the equilibrium and of the mechanically excited state of the star seem in contradiction, they can be justified as follows.  The hydrostatic equilibrium is appropriate if a liquid description  holds. This is a reasonable approximation in any layer of the strange star. As is well known, for matter density above $10^5$ g/cm$^{3}$ the pressure in the ionic crust is mainly due to degenerate electrons that can be treated as a fluid. The remaining part of the ionic crust, having smaller density, has a very small mass that can be safely neglected in the solution of the hydrostatic equilibrium equations. Regarding quark matter, the CFL phase is a fluid and thus a liquid description is appropriate. The CCSC phase is rigid because of the rigidity of the  periodic pattern of the gap parameter, however the equilibrium pressure of quark matter does weakly depend on the presence of the condensate. We shall assume that such a contribution can be absorbed in a liquid-like description of quark matter. 

Regarding the mechanical response, we will  only include the effect of the shear modulus, because we focus on torsional oscillations that only have a tangential stress component. Obviously, these oscillations are confined to the crusts of the nonbare strange star, because a fluid cannot support shear waves. The  various interfaces between the different layers  will  be described by  appropriate boundary conditions. 
Assuming that a small shear perturbation acts on the system,  we expand the displacement field, the  pressure and the matter density respectively as follows
\be
\bm  U=  \bm  U_0 + \bm u \,\,\,\,\,\, P_{ik} = P_0 \delta_{ik} - \Pi_{ik} \,\,\,\,\,\, \rho= \rho_0 + \delta\rho\,,  \label{eq:fluctUP}
\ee
where the quantities with a subscript $0$ correspond to  the time independent background and the remaining quantities represent the linear perturbations.

\section{Background configuration}\label{sec:background}
We set as the equilibrium configuration the one  with $\bm  U_0=0$ and assume negligible background magnetic field. Therefore, our analysis is strictly valid for slowly rotating  nonmagnetar stars; however we will briefly   comment on the effect of a nonvanishing magnetic field.  In the stationary state, electrons and quarks are confined in regions with a net charge and are in thermal equilibrium. This is possible because nonneutral systems in appropriate geometries can be confined and in thermodynamic equilibrium; a typical example are non-neutral plasmas, see~\cite{Dubin:1999zz} for a review, which, unlike quasineutral plasmas, can be brought to equilibrium and confined. 

\begin{figure}[t!]
\includegraphics[width=8.cm]{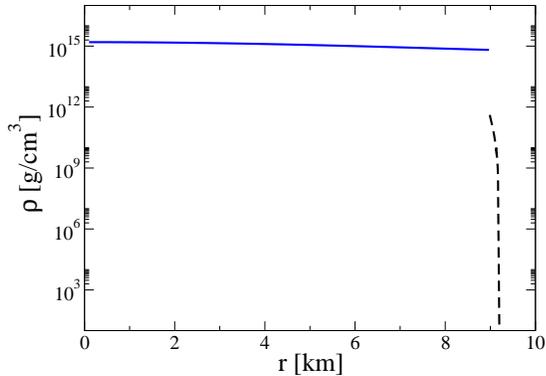} 
\caption{Density profile of the considered model of nonbare strange star. Strange matter is present for  $r <R_q\simeq 8.98$ km; the solid (blue) line represents the quark matter density.  The ionic crust layer extends for  $R_q < r <R$, with $R=9.18$ km  the star radius;  the dashed (black) line represents the nuclear matter density.
 \label{fig:rho}}
\end{figure}

The equilibrium is determined by the balance between the hydrostatic pressure and the gravitational attraction, which is appropriately described by  the   Tolman-Oppenheimer-Volkov (TOV) equation
\be
\frac{\partial p}{\partial r}=
-\frac{G (p+ \rho ) \left(m+4 \pi  p \, r^3\right)}{r (r-2 G \, 
   m)} \,,
\label{eq:tov}
\ee
where  $m(r) = \int^r_0 dr' \, r'^2 \, \rho(r')$ and $G$ is the gravitational constant. Two different EoSs must be used, $p_\text{QM}(\rho)$, for quark matter in the quarksphere and  $p_\text{NM}(\rho)$, for nuclear matter in the ionic crust.    
In both cases we  assume that the temperature is so low that the background  distribution can be approximated by a Fermi liquid at zero temperature. Quarks are strongly interacting and a first principle calculation of the EoS is unfeasible. However, if the leading contribution of the quark interaction is quark pairing, we expect corrections of the order $\Delta^2/\mu^2 \lesssim 10 \% $ to the free Fermi  gas EoS.
To effectively take into account the strong interaction we use  the general parameterization of the quark matter EoS  given in \cite{Alford:2004pf}
\be
\Omega_{\text{QM}} = -\frac{3}{4 \pi^2} a_4 \mu^4 + \frac{3}{4 \pi^2}  a_2 \mu^2 + B_{\text{eff}}\,,
\ee\label{eq:EoS_quark_matter}
where $a_4$, $a_2$ and $B_{\text{eff}}$ are independent of the average quark chemical potential $\mu$. We use the set of parameters $a_4=0.7$, $a_2=(200$ MeV)$^2$ and $B_{\text{eff}}=(165$ MeV)$^4$  (similar results hold for the two sets of parameters  discussed in~\cite{Mannarelli:2014ija}).

For the ionic crust EoS we assume that it consists of a Coulomb crystal embedded in a degenerate electron gas. We only need an expression at density below the neutron drip point, thus we use the data reported in~\cite{Haensel:1993zw}, see~\cite{Datta:1995}  for various nuclear matter EoSs. 
As in~\cite{1992ApJ...400..647G} we assume that the highest density of nuclear matter corresponds to the neutron drip point.  This assumption corresponds to the maximization of the 
ionic crust radius. More refined studies show that properly taking into account the  electronic pressure  results in a reduction of the ionic crust mass and thickness~\citep{Martemyanov:1994ly,1997ChPhL..14..314H}. However, the detailed extension of the ionic crust is not relevant for our purposes, we are only interested in order of magnitude estimates of the effect of a ionic crust.
Therefore, in our approach the radius, $R_q$, corresponding to the boundary between the quarksphere and the ionic crust, is determined by the condition that the pressure of quark matter equals the pressure of nuclear matter at the neutron drip point, {\it i.e.} $p_\text{QM}(R_q) = p_\text{NM}(R_q) = p_\text{ND} \simeq 7.8 \times 10^{29}$ dynes/cm$^{2}$.  The star radius, $R$, is determined by the  boundary condition on the pressure $p_\text{NM}(R) =0$,  meaning that the pressure at the surface of the ionic crust vanishes. 
Note that the effect of the magnetic field on the matter distribution is small even considering the extreme values characteristics of magnetars, see for example the discussion in~\cite{Frieben:2012dz}. 

In Fig.~\ref{fig:rho} we report the matter density profile for the considered strange star model obtained solving the TOV equation for a strange star having mass of about $1.4$ M$_\odot$. The obtained star radius is $R \simeq 9.18$ km, corresponding to the outer surface of the ionic crust layer. The matter density in the quarksphere is roughly constant, changing by less than a factor $2$. There is a jump at $R=R_q \simeq 8.98 $ km  corresponding to the transition between quark matter and the ionic crust. The nuclear density changes by about $10$ orders of magnitude across the ionic crust layer, from the iron-like density of $\rho_\text{Fe} \simeq 7.8$ g/cm$^{3}$ to the neutron drip density $\rho_\text{ND} \simeq 4.3\times 10^{11}$ g/cm$^{3}$. This is the same radial dependence found in the outer crust of standard neutron stars.

\section{Torsional Oscillations}\label{sec:torsional}
We now discuss the fluctuations of the displacement fields. Assuming a fluid description, the oscillations obey the Euler's equation
\be\label{eq:euler_a}
\rho \frac{d u_i}{d t} +  \partial_k \Pi_{ik}  = 0\,,
\ee 
and the continuity equation  
\be\label{eq:continuity_a}
\partial_t \delta\rho + \bm \nabla \cdot (\rho\, \bm u) = 0\,.
\ee
For simplicity we have not considered dissipative processes and assumed that  the matter density of the various species does not change, in other words  the system is in ``chemical" equilibrium. We have also omitted the collision term, meaning that we assume that collisions are so fast that local equilibrium is reached  in a time scale negligible compared with the one of external forces. All the perturbations oscillate at the same frequency $\sigma$; in particular we focus on the  eigenmodes
\be
\bm u = e^{i \sigma t} \bm \xi\,,
\ee
and assume  that all of the fluctuations are proportional to the displacement field, meaning that any perturbation vanishes for $\bm \xi=0$. 
We  restrict the analysis to  torsional oscillations,  corresponding to transverse oscillations with no radial displacement
\be\label{eq:torsional}
\bm \nabla \cdot \bm u =0 \qquad u_r=0.
\ee
Upon plugging the above expressions in Eq.~\eqref{eq:continuity_a}  it can be easily shown that no matter density fluctuation is produced.

For nonrotating and nonmagnetic stellar models the torsional oscillations do not couple with any other star oscillation, thus  an eigenmode analysis is possible. According with~\cite{1988ApJ...325..725M} we will indicate with $_\ell t_n$ the torsional mode having harmonic index $\ell$ and $n$ nodes. The $t$-mode displacement field can be decomposed  as follows
\begin{align}\label{eq:defW}
\xi_\theta &= W \frac{1}{\sin \theta} \frac{\partial Y_{lm}}{\partial \phi}  \qquad \xi_\phi = -W \frac{\partial Y_{lm}}{\partial \theta}\,,
\end{align}
and in the Newtonian approximation (see \cite{1983MNRAS.203..457S, Andersson:2002jd}  for general relativistic discussions)  the Euler's equation  in spherical coordinates reads
\be
\begin{split}
\sigma^2 W_i= & v_i^2 \left[ -\frac{d \log \nu_i}{d r} \left(\frac{d W_i}{d r} - \frac{W_i}{r} \right) \right. \\ &-\left. \frac{1}{r^2}\frac{d}{d r} \left( r^2\frac{d W_i}{d r}  \right) +\frac{\ell (\ell +1)}{r^2} W_i \right]\,,
\label{eq:Wi}
\end{split}\ee
where $v_i$ is the shear velocity and $\nu_i$ is the shear modulus. The index $i=1,2$ refer to the CCSC crust layer and to the ionic crust layer, respectively.  The differential equations describing the oscillations of the two crusts are decoupled, because no long range forces are present. However, a coupling between the two oscillations is produced by the  boundary conditions (BCs), effectively describing the effect of short-range electromagnetic forces between the two crusts. The set of BCs that we use for  solving the two Euler's equations is the following
\begin{align}
\left.\left(\frac{d W_1}{d r} - \frac{W_1}{r} \right)\right |_{R_\text{CFL}} &= 0 \label{eq:CFL-CCSC}\\
\nu_1\left.\left(\frac{d W_1}{d r} - \frac{W_1}{r} \right)\right |_{ R_\text{q}} &=\nu_2 \left.\left(\frac{d W_2}{d r} - \frac{W_2}{r} \right)\right |_{R_\text{q}}\label{eq:CSC-ionic_NT}\\
W_1(R_\text{q}) &= W_2(R_\text{q})\label{eq:CSC-ionic_NS}\\
\left.\left(\frac{d W_2}{d r} - \frac{W_2}{r} \right)\right |_{R_2} &= 0\,, \label{eq:ionic-vacuum} 
\end{align}
where $R_2\simeq 9.15$ km corresponds to the boundary between the solid crust and the ocean (at $\rho \simeq 10^7 $g/cm$^{3}$).  These equations describe  the no-traction BCs, Eqs.~\eqref{eq:CFL-CCSC}, \eqref{eq:CSC-ionic_NT}, \eqref{eq:ionic-vacuum}, and  the no-slip BC, Eq.~\eqref{eq:CSC-ionic_NS}. The no-traction BCs simply mean that no net force acts on any boundary between two adjacent layers. The no-slip BC means that the displacement of the two crust layers at the interface is the same. Thus, we are assuming that the static friction at this interface is so strong that at any time the two crust oscillations  have the same amplitude and frequency, $\sigma$. Note that the assumption that the oscillations in the two crusts have the same phase and frequency was implicitly done in Eq.~\eqref{eq:Wi}.

At the CCSC-CFL boundary we only impose the no-traction BC, Eq.~\eqref{eq:CFL-CCSC}, but the two materials have free slip. The reason is that the CFL phase is a neutral superfluid, therefore it should not react with a force to the oscillation of the CCSC internal surface.  In this case we are assuming that no vortices are created by the oscillation of the CCSC boundary and that the temperature is so low that thermal fluctuations in the CFL superfluid can be neglected. A similar reasoning applies to the ionic crust-ocean boundary, which is described by the no-traction BC, Eq.~\eqref{eq:CSC-ionic_NS} and free slip.

The shear moduli and densities of the two crust layers are  fundamental ingredients for determining the  solutions of Eq.~\eqref{eq:Wi}. We already determined the matter density profile in Sec.~\ref{sec:background}. Regarding the shear modulus of 
the CCSC phase, it can be obtained from  the low-energy Lagrangian description in terms of  phonon-like excitations, see~\cite{Casalbuoni:2001gt,Casalbuoni:2002pa,Casalbuoni:2002my,Mannarelli:2007bs}.  In particular,  the coefficients of the quadratic terms of the low-energy Lagrangian  determine the  linear response of the crystal to excitations. Unfortunately, to extract the shear modulus various approximations have to be used: the procedure used in~\cite{Mannarelli:2007bs}  relies on a  Ginzburg-Landau (GL)  expansion of the Nambu-Jona Lasinio model used to mimic the properties of quark matter at the relevant baryonic densities. It is known that this procedure is not under quantitative control, see~\cite{Mannarelli:2006fy}, moreover the recent analyses of two flavor systems~\citep{Cao:2015rea} seem to indicate that many terms in the GL expansion must be included to have a controlled approximation scheme. Using the  GL expansion up to order $\Delta^2$ it was found in~\cite{Mannarelli:2007bs}  that 
\be
\label{eq:nu}
\nu_\text{CCSC} \simeq \nu_0\left(\frac{\Delta(\mu)}{10 \text{ MeV}}\right)^2\left(\frac{\mu}{400 \text{ MeV}}\right)^2\,,
\ee
where
\be \label{eq:nu0}
\nu_0= 2.47 \frac{\text{MeV}}{\text{fm}^3}\,, 
\ee
will be  our reference value.  We will  assume that the shear modulus is constant within the CCSC crust and hereafter we will  identify the CCSC shear modulus with the reference value. It is certainly true that  $\nu_\text{CCSC}$ depends on the quark chemical potential $\mu$. However, in the interior of compact stars the quark chemical potential is almost constant.  Moreover, as we will  see below, the most relevant aspect is that  $\nu_\text{CCSC}$ is much larger than the  shear modulus of the ionic crust. Regarding the quark matter density, it does not strongly depend on the radial coordinate. Therefore, we  will  consider the constant value $\rho_\text{QM} = 10^{15}$ g/cm$^{3}$; a typical quark matter value, see Fig.~\ref{fig:rho}. 

The shear modulus of the ionic crust depends on the particular crystalline structure considered and on the plane of  application of the shear stress. Early calculation of the shear modulus of monovalent crystals were performed by~\cite{1936RSPSA.153..622F}. In compact stars the orientation of the crystals is unknown,  however it is possible to define an effective shear modulus, $\nu_\text{eff}$,  averaging over directions,  as shown in~\cite{Strohmayer}, which should give an excellent approximate result if the crust has a polycrystalline structure. From dimensional analysis and considering that the rigidity is due to the electromagnetic interaction between ions, it is clear that $\nu_\text{eff}(r)  \propto (Z(r)e)^2 n_N(r)^{4/3} $, where $Z(r)$ is the radial dependent proton number and $n_N(r)$ is the number density of nuclei. The order of magnitude estimate of the shear modulus gives a  very large value, of about $10^{30}$ dyne cm$^{-2}$ for the inner crust, as already noted in~\cite{Smoluchowski:1970zz}. Using Monte Carlo simulations, see~\cite{Strohmayer}, or Molecular Dynamics methods, see~\cite{Hoffman:2012nr}, eventually including quantum fluctuations as in~\cite{2011MNRAS.416...22B}, one can estimate the proportionality factor,  finding that 
\be\label{eq:nueff}
\nu_\text{eff}(r) = c \frac{n_N(r) (Z(r)e)^2}{a(r)} \,,
\ee
where  $a(r)=(3/(4 \pi n_N(r)))^{1/3}$ is the average inter-ion spacing  and  $c \sim 0.1$ gives an approximate result at any density larger than $10^6$ g/cm$^{3}$, where electrons form a degenerate Fermi gas and the Coulomb crystal model can be applied. At smaller density, the Coulomb crystal model cannot be applied and Eq.~\eqref{eq:nueff} does not apply. As an example,  extrapolating Eq.~\eqref{eq:nueff} to the star surface does not reproduce the known shear modulus of iron.   

Summarizing,  the matter densities and shear moduli of the nonbare strange star are approximately given by 
\be\label{eq:nu-toy}
\nu = \left\{  \begin{array}{ll}  0 &\qquad\text{for   } r< R_\text{CFL}\\
\nu_0 &\qquad\text{for   }R_\text{CFL} <r <R_\text{q} \\
\nu_\text{eff}(r) &\qquad\text{for   }   R_\text{q} <r <R_2
\end{array} \right.
\ee

and

\be\label{eq:rho.mu-toy}
\rho = \left\{  \begin{array}{ll}  
\rho_\text{QM} &\text{for }  R_\text{CFL} <r <R_\text{q}  \\
\rho_\text{NM}(r) &\text{for }  R_\text{q} <r <R_2
\end{array} \right.
\ee
with  $\rho_\text{QM} = 10^{15} $ g/cm$^{3}$ and the CCSC shear modulus given in Eq.~\eqref{eq:nu0}. For a realistic description of amplitude and frequency of the torsional oscillations, the radial dependence of the nuclear matter density and of the effective shear modulus must be appropriately taken into account in Eq.~\eqref{eq:Wi}. However, to disentangle the various aspects of the problem in Sec.~\ref{sec:two_homo}  we will  approximate the ionic crust as an homogeneous system. We turn to a discussion of the inhomogenous ionic crust layer in Sec.~\ref{subsec:inho}.

\subsection{System of two homogeneous crusts}\label{sec:two_homo}
Let us approximate the nonbare strange star crust as made by two homogenous crust layers. We consider two different parameter sets for characterizing the properties of the ionic crust. Parameter set A  corresponds to the nuclear matter density and the effective shear modulus at the inner surface of the ionic crust, respectively given by
$\rho_\text{NM}(R_q) = \rho_\text{ND} $ and $\nu_\text{eff} (R_q)\simeq 3.4\cdot 10^{-5}$MeV/fm$^3$. The parameter set B corresponds to the nuclear matter density  and the effective shear modulus near the ocean surface, respectively given by $\rho (R_2) =2.8\cdot 10^9 $g/cm$^{3}$ and $\nu_\text{eff} (R_2) = 2.1\cdot 10^{-10} $MeV/fm$^3$.

Since both crust layers are homogeneous,  Eq.~\eqref{eq:Wi} can be cast in the form of  two Bessel's equations, and  we can determine an analytic expression for the oscillations for any $\ell$. In the following we focus on  $\ell=1$, but similar results hold for different values of $\ell$.  Imposing the BCs given in Eq.~\eqref{eq:CFL-CCSC} and  \eqref{eq:ionic-vacuum} we find that
\be
W_i(r) = \frac{1}{\sqrt{r}} C_i \left( J_{3/2} (r \sigma/v_i) -  K_i Y_{3/2} (r \sigma/v_i)   \right)\,,
\ee 
where $J_n$ and $Y_n$ are Bessel functions of the first and second kind, respectively, and \be K_1=J_{5/2} (a R_1 \sigma/v_1)/Y_{5/2} (a R_1 \sigma/v_1) \ee and  \be K_2=J_{5/2} (R \sigma/v_2)/Y_{5/2} (R \sigma/v_2)\,. \ee By the two remaining BCs, Eq.~\eqref{eq:CSC-ionic_NT} and Eq.~\eqref{eq:CSC-ionic_NS}, we can determine the ratio $C_1/C_2$ and the eigenmode frequency.

\begin{figure}[t!]
\includegraphics[width=8cm]{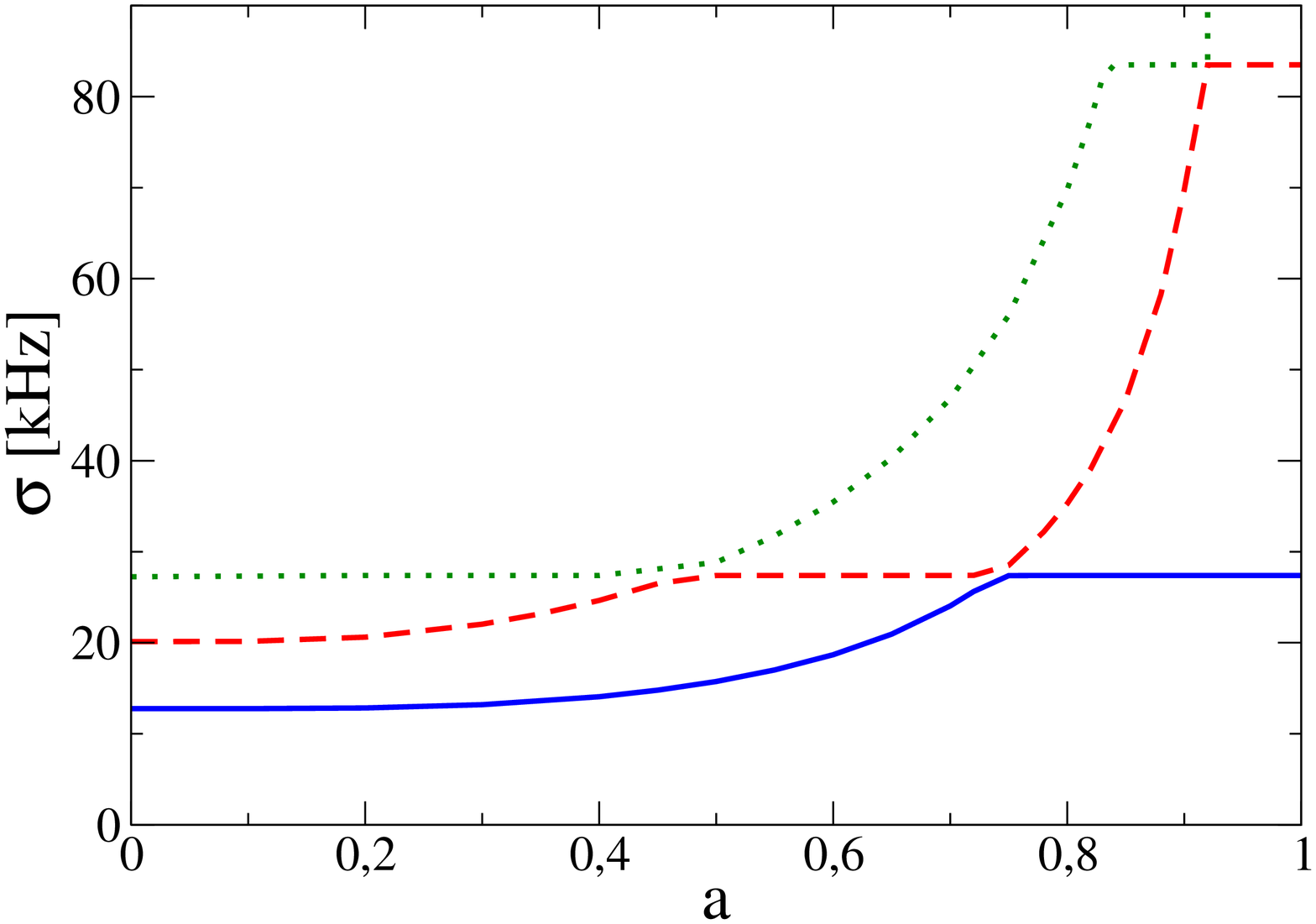}   
\includegraphics[width=8cm]{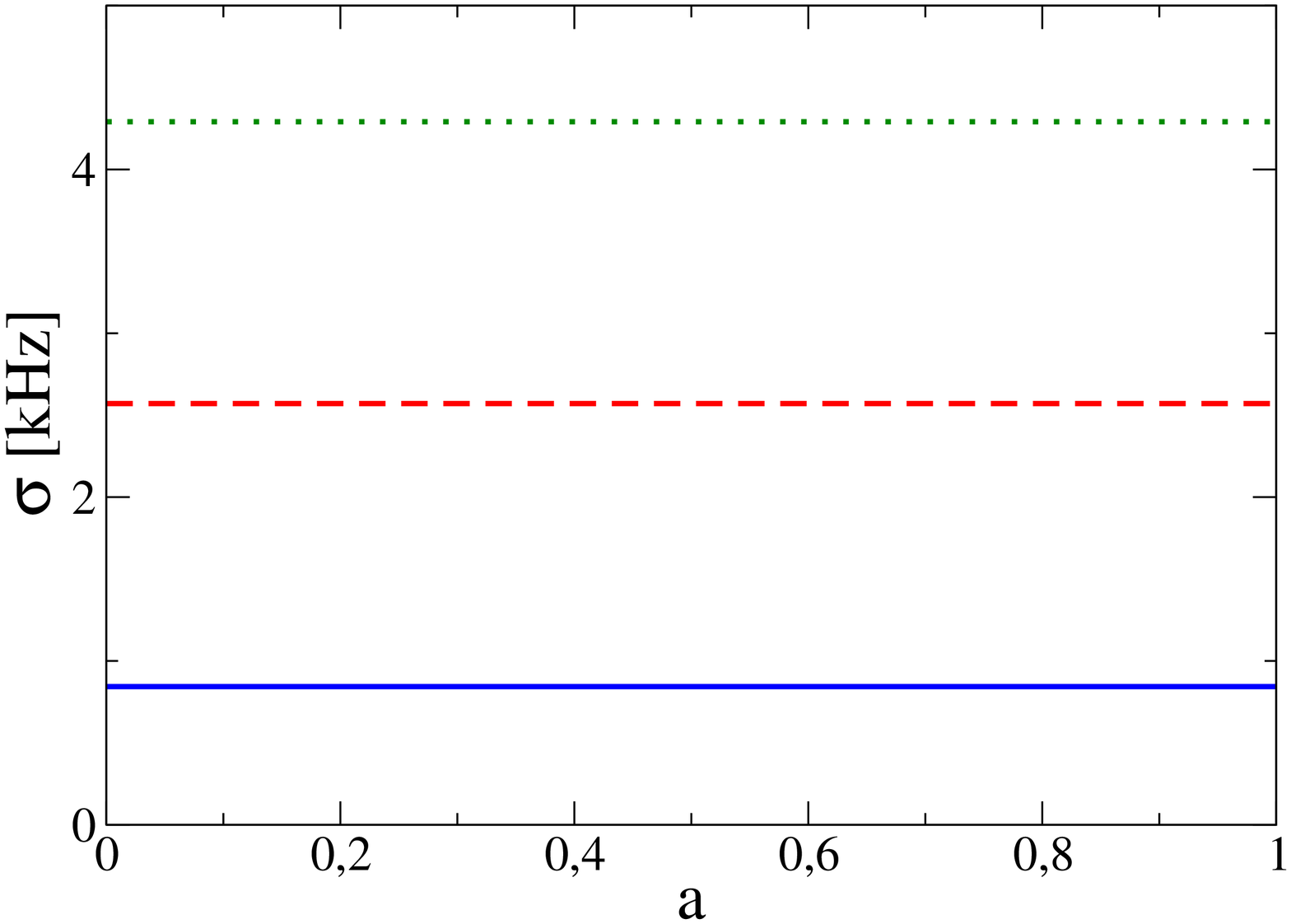} 
\caption{Frequency of the $t$-modes with $\ell=1$ as a function of $a$ for different number of nodes:  $n=1$ mode solid blue, $n=2$ mode dashed red,  $n=3$ mode dotted green.  The plots have been obtained  considering both the CCSC crust layer and the ionic crust layer as  homogeneous.  For the ionic crust we considered two sets of parameters. 
Top: $v_2\simeq 3.55\times 10^8$ cm/s and $\rho=\rho_\text{ND}$.  The frequency of the $_1t_1$ mode increases with $a$ until $a\simeq 0.75$. For $a\gtrsim 0.75$  this mode is an almost pure ionic-crust oscillation and does not depend on the extension of the CCSC crust layer. A similar phenomenon happens for the $_1t_2$ mode, which becomes an almost pure   ionic-crust oscillation for $0.5 \lesssim a \lesssim 0.75$, and for the $_1t_3$ mode, which becomes an almost pure ionic-crust oscillation for $0.85 \lesssim a \lesssim 0.9$.
Right panel: $v_2\simeq 1.1\times 10^7$ cm/s and $\rho \simeq 1.5$ g/cm$^3$. The frequencies of the  modes $_1t_1$, $_1t_2$ and $_1t_3$ are all very weakly dependent on $a$. 
\label{fig:frequencies_cost}}
\end{figure}

\begin{figure}[t!]
\includegraphics[width=8cm]{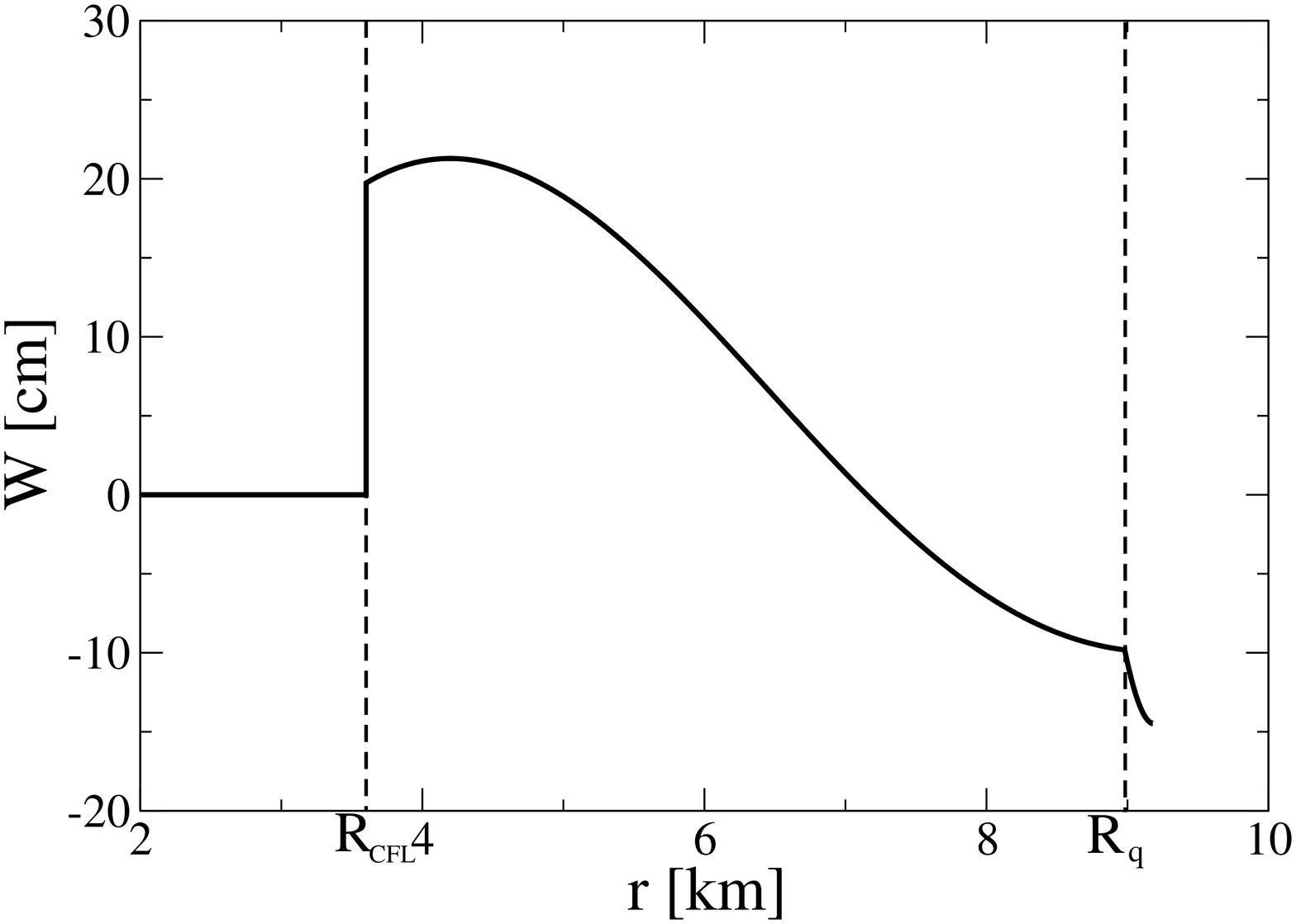}
\includegraphics[width=8cm]{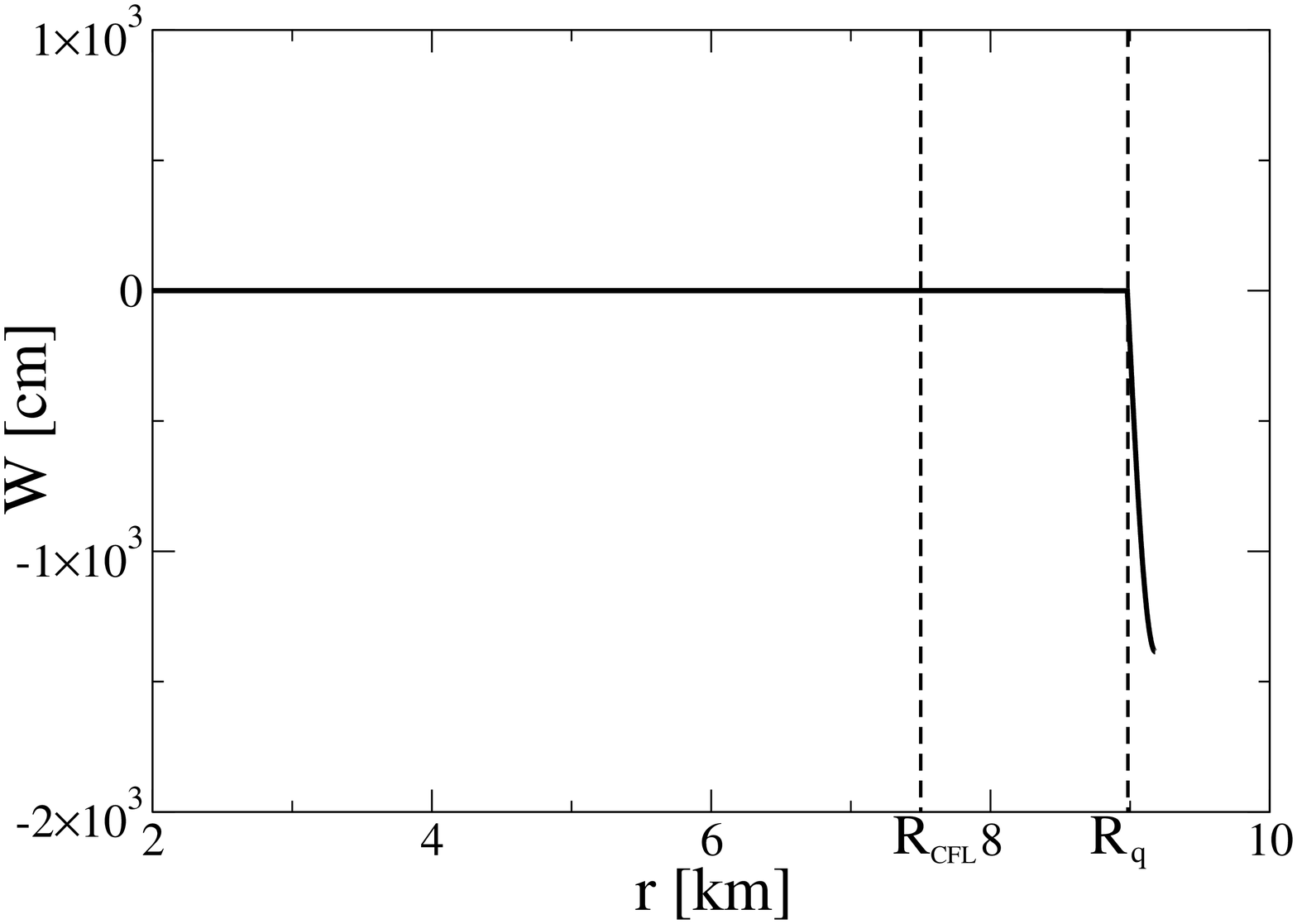}
\caption{Amplitude of the oscillation of the $_1t_1$ mode for the parameter set A for two different values of $a$. 
Top: results obtained  for $a=0.4$. Both crust layers oscillate with a comparable amplitude. Bottom: results obtained for $a=0.8$. The oscillation is  basically segregated to the ionic crust. In both cases the amplitudes are obtained assuming that the energy of a  Vela-like glitch is conveyed to  the $_1t_1$ mode oscillation. 
\label{fig:AmpcasA}}
\end{figure}

In Fig.~\ref{fig:frequencies_cost} we report the dependence of the frequency of the first three eigenmodes on $a$. The solid blue line corresponds to the  $_1t_1$ mode; the dashed red line corresponds to the $_1t_2$ mode  and the dotted green line corresponds to the  $_1t_3$ mode. The results reported in the upper panel are obtained with the parameter set A, while the results reported in the bottom panel are obtained with the parameter set B.

Let us focus on the fundamental  $_1t_1$ mode obtained with the parameter set A. For $a\lesssim 0.75$ the frequency of this mode increases with increasing $a$. Then, for $a\gtrsim 0.75$ the frequency becomes almost independent of $a$. 
For  $a\lesssim 0.75$ the frequency and the amplitude of the oscillation are very similar to the results obtained in~\cite{Mannarelli:2014ija} for a bare strange star. In this range of values of $a$ the system behaves as if the ionic crust layer is absent. For $a\gtrsim 0.75$ the opposite behavior happens, with torsional oscillation segregated in the ionic crust layer with a frequency determined by the mechanical properties of the ionic lattice. Indeed, this frequency corresponds to $\omega_2 \simeq \pi/2\cdot v_2/(R-R_q)$ as described in the Appendix \ref{sec:rigid_slab} using a simpler model with planar geometry. For $a^*=[1-2 v_1/R_q\cdot (R-R_q)/v_2] \sim 0.75$  there is a crossing between two different types of mode oscillations, meaning that the frequency of the first CCSC crust oscillation and of the first ionic crust oscillation are equal. Therefore, at this point there is a crossing between the frequency of the $n=1$ mode and of the $n=2$ mode. The transition between the two behaviors corresponds to a node at the interface between the two crust layers.

The modes with more nodes have a similar behavior. For the $_1t_2$  mode there are two crossing points, one with the mode with $n=3$ and one with the mode with $n=1$. As can be seen in the upper panel of Fig.~\ref{fig:frequencies_cost}
 there exist two regions in which the frequency of  $_1t_2$  mode is almost independent of $a$ and two regions in which it is strongly dependent on $a$. The  $_1t_3$ mode has a similar behavior, as well. We have checked that for any considered mode,  the range of values for which the frequency is almost constant corresponds to  torsional oscillation basically confined to the ionic crust layer. Note that for any value of $a$ there is at least one of the three modes that is weakly dependent on $a$, meaning that there is at least one mode that is segregated to the ionic crust.  
For the set of parameter B,  the frequencies of the three considered modes are independent of $a$ for any value of $a$, see the bottom panel of Fig.~\ref{fig:frequencies_cost}. Therefore, for the parameter set B the low lying $t$-mode frequencies are  completely determined by the mechanical properties of the ionic crust layer.

To better understand the difference between the two regions of frequencies, respectively dependent and constant in $a$, let us discuss in detail  the  amplitude of the $_1t_1$ mode. In Fig.~\ref{fig:AmpcasA} we show the amplitude of the $_1t_1$ mode obtained for the parameter set A, considering $a=0.4$ upper panel, and $a=0.8$, bottom panel. To determine the absolute values of this amplitude we assume that all the energy of a  Vela-like glitch, $E_\text{Vela}\sim5\cdot10^{42}$ergs, is conveyed to the  $_1t_1$ mode. If a fraction $\alpha$ of the  Vela-like glitch is considered, then the amplitude has to be scaled by $\sqrt{\alpha}$. For $a=0.4$, we obtain that the CCSC crust and the ionic crust oscillate with similar amplitude that is order of tens of centimeters. For $a=0.8$ the oscillation of the CCSC crust is negligible and the mode is segregated to the ionic crust layer. The energy of the glitch conveyed in the thin ionic layer generates amplitude order of tens of meters. This result confirms the expectation that $t$-modes having an $a$ independent frequency are basically ionic crust oscillation. We can therefore classify the nonbare strange star oscillations as CCSC crust oscillations and ionic crust oscillations.    
For the set of parameters B the dependence from $a$ of the first three modes is  negligible and the corresponding amplitude is always confined in the ionic crust layer with a  behavior very similar to the one reported in the bottom panel of Fig.~\ref{fig:AmpcasA}. However, the absolute value of the amplitude in this case is extremely large and can reach values of kilometers at the star surface. Since these values of the oscillation amplitude are well beyond the linear approximation region, let us understand the physical behavior of the system  considering the deformation of the crust.

\begin{figure}[t!]
\includegraphics[width=8cm]{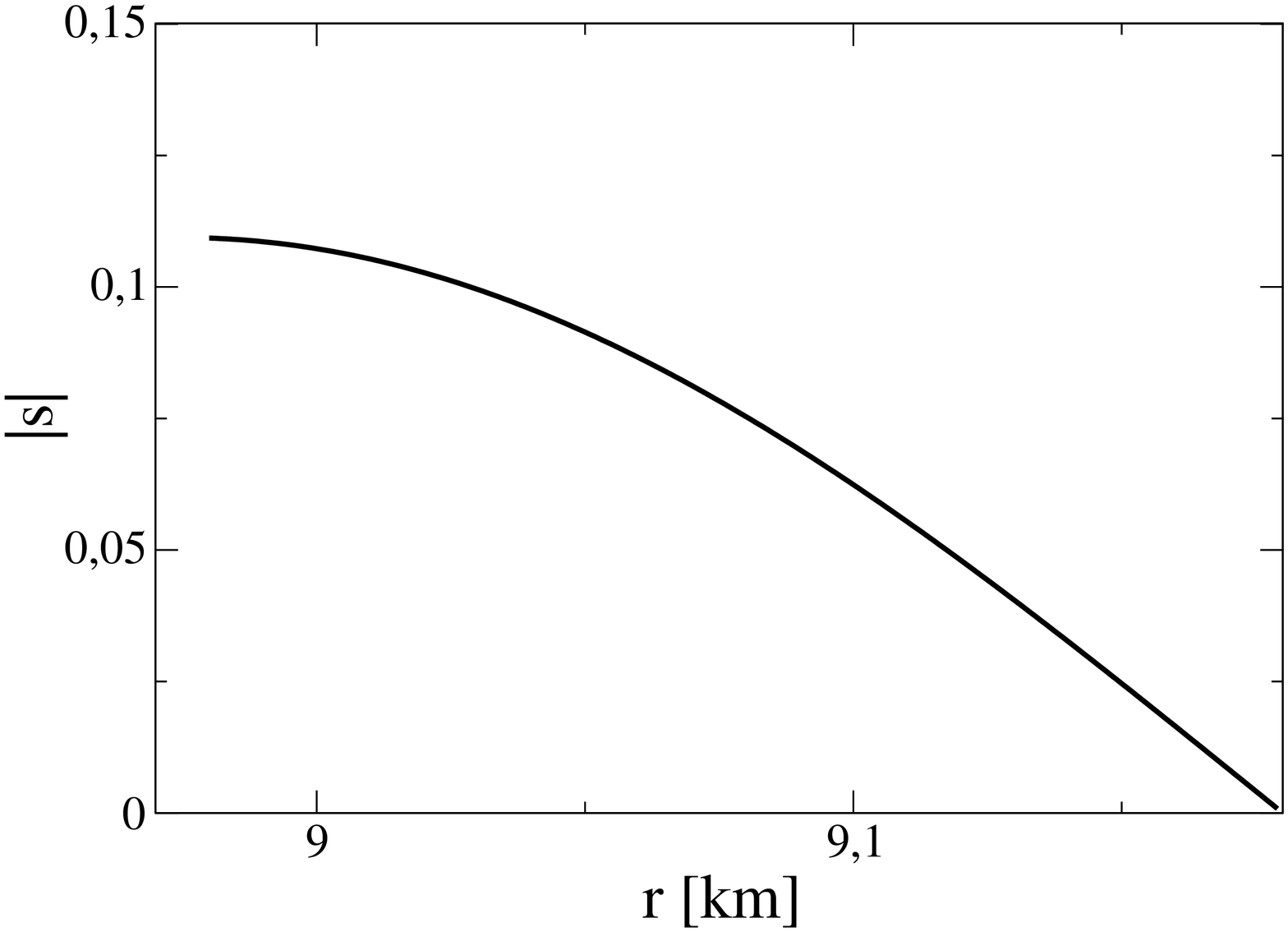} 
\includegraphics[width=8cm]{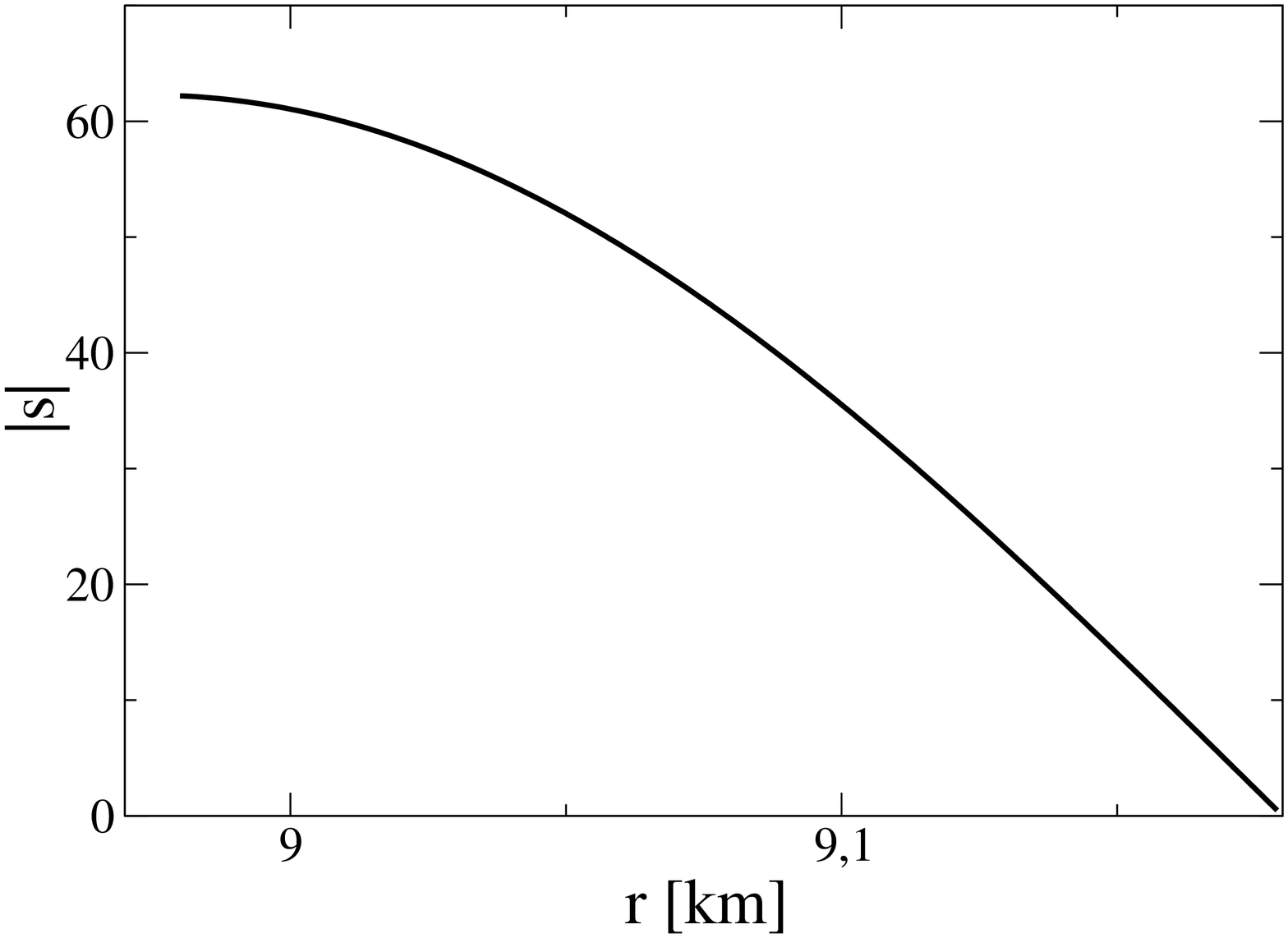} 
\caption{Absolute values of the shear strain, defined in Eq.~\eqref{eq:abss}, across the ionic crust layer.  Top:  results obtained with parameter set A and for $a=0.8$. Bottom:  results obtained with parameter set B and for $a=0.4$. In both cases the strains are obtained assuming that the energy of a  Vela-like glitch is conveyed into a $_1 t_1$ mode oscillation.  
\label{fig:stain_cost}}
\end{figure}

A measure of a solid deformation is the shear strain (also called the shear  traction). 
The shear strain depends on the angular coordinates, but for simplicity we define  
\be\label{eq:abss} |s|=\left|\frac{dW}{dr}-\frac{W}{r}\right| \,,\ee which only depends on the radial coordinate;  the shear strain at any angle can be obtained multiplying $|s|$ by the appropriate angular function, see Eq.~\eqref{eq:defW}. We focus on the $_1t_1$ mode and in Fig.~\ref{fig:stain_cost}
we report the corresponding shear strain across the  ionic crust (the deformation of the  CCSC crust layer is always much smaller). The upper panel of Fig.~\ref{fig:stain_cost} corresponds to the ionic crust strain produced by a  $_1t_1$ ionic crust oscillation with the parameters set A and $a=0.8$, namely the reported  strain corresponds to the amplitude shown in upper panel of Fig.~\ref{fig:AmpcasA}. The bottom panel of  Fig.~\ref{fig:stain_cost} corresponds to a ionic crust strain induced by a $_1t_1$  oscillation for the parameters set B, showing the extreme values for the deformation discussed above. Note that in both panels of Fig.~\ref{fig:stain_cost} the strain has a maximum at the boundary between the CCSC crust layer and the ionic-crust layer. The shear strain is a monotonic decreasing function of the radial coordinate because the ionic crust layer is homogenous and because the shear strain at the  surface has to vanish because of the no-traction BC. As we will see in the next section, considering an inhomogenous matter density results in a nonmonotonic shear strain. For all considered cases the deformation of the ionic crust layer is large meaning that the linear approximation is not valid. Indeed, in this case the crust  very likely cracks  before reaching such a large deformation. We will discuss  crust cracking in Sec.~\ref{sec:conclusions}, we now turn to a more realistic description of the ionic crust layer.

\subsection{Inhomogeneous  ionic crust}\label{subsec:inho}
\begin{figure}[h!]
\includegraphics[width=8cm]{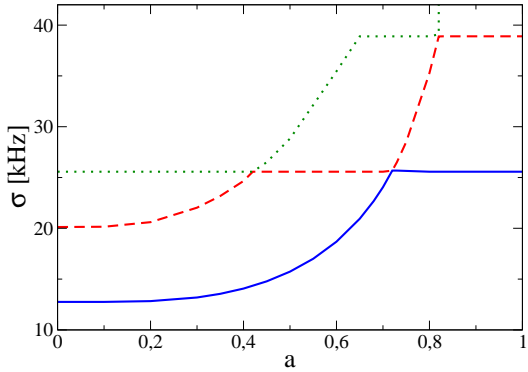} 
\caption{Frequency of the $t$-modes with $\ell=1$ as a function of $a$ for different number of nodes:  $n=1$ mode solid blue, $n=2$ mode dashed red,  $n=3$ mode dotted green. These results have been obtained  considering  the CCSC crust layer homogenous and the radial dependence of the shear modulus and of the matter density of the ionic crust layer. \label{fig:frequencies_noncost}} 
\end{figure}

We now include the radial dependence of the mechanical properties of the ionic crust matter, while keeping both the matter density of quark matter density  and the shear modulus of the CCSC crust layer constant. In particular, we numerically solve  Eq.~\eqref{eq:Wi} including the radial dependence of the ionic crust shear velocity, $v_2(r)$,  and shear modulus, $\nu_2(r)$. The frequencies of the first three torsional eigenmodes are reported in Fig.~\ref{fig:frequencies_noncost}. This figure is rather similar to the one reported in the upper panel of Fig.~\ref{fig:frequencies_cost} showing that the parameter set A is the one that gives a reasonably good approximation of the ionic crust, as far as $t$-mode oscillation are concerned.

\begin{figure}[h!]
\includegraphics[width=8cm]{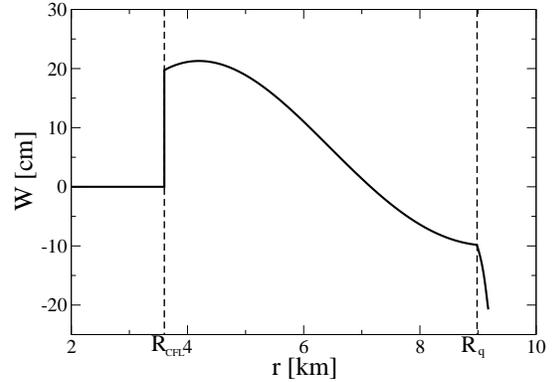} 
\includegraphics[width=8cm]{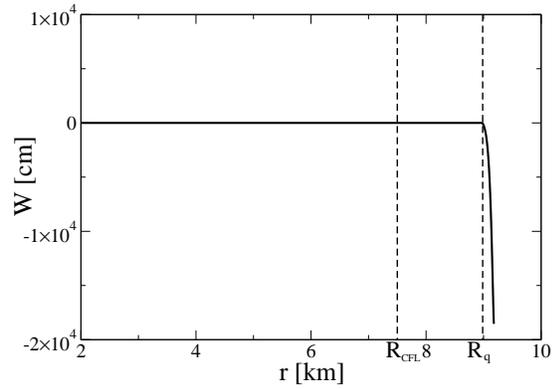} 
\caption{Top: Amplitude of the oscillation of the fundamental mode for $a=0.4$. Both crust layers oscillate with a comparable amplitude.  Bottom: Amplitude of the oscillation of the fundamental mode for $a=0.8$. The oscillation is basically segregated to the ionic crust. \label{fig:oscillations}}
\end{figure}

Let us focus on the $_1t_1$ mode. For small values of $a$ the CCSC crust oscillation frequency strongly depends on $a$. The corresponding oscillation amplitude is mostly confined in the CCSC crust, see the upper panel of Fig.~\ref{fig:oscillations} for a representative behavior obtained with $a=0.4$. This amplitude is very similar to the one reported in the upper panel of  Fig.~\ref{fig:AmpcasA}; the reason is that the same values of the CCSC crust density and shear modulus have been used. Increasing the value of $a$ one reaches a critical value around $a^{*}\sim 0.7$ for which there is a crossing between the $n=1$ and the $n=2$ modes.  For $a>a^{*}$ the amplitude of the oscillations is mainly confined in the ionic crust layer and the oscillation frequency is almost independent of $a$. In the bottom panel of Fig.~\ref{fig:oscillations} is reported the amplitude of the  $_1t_1$  ionic-crust oscillation obtained for $a=0.8$. In this case, the qualitative  behavior of the oscillation amplitude is very similar to the one reported in the bottom panel of Fig.~\ref{fig:AmpcasA}, but the amplitude of  the oscillation is now larger by about an order of magnitude. The reason is that in this case the light matter at the top of the ionic crust is easily displaced by the torsional oscillation. This effect is akin to the seismic site effect leading to the amplification of earthquake seismic waves  in presence of certain geological conditions. In other words, considering a constant density one underestimates the matter displacement.

As in the  case of an homogenous ionic crust considered in the previous section, the amplitude of the oscillations is very large and indicates that a  crust cracking may occur.

\begin{figure}[h!]
\includegraphics[width=8cm]{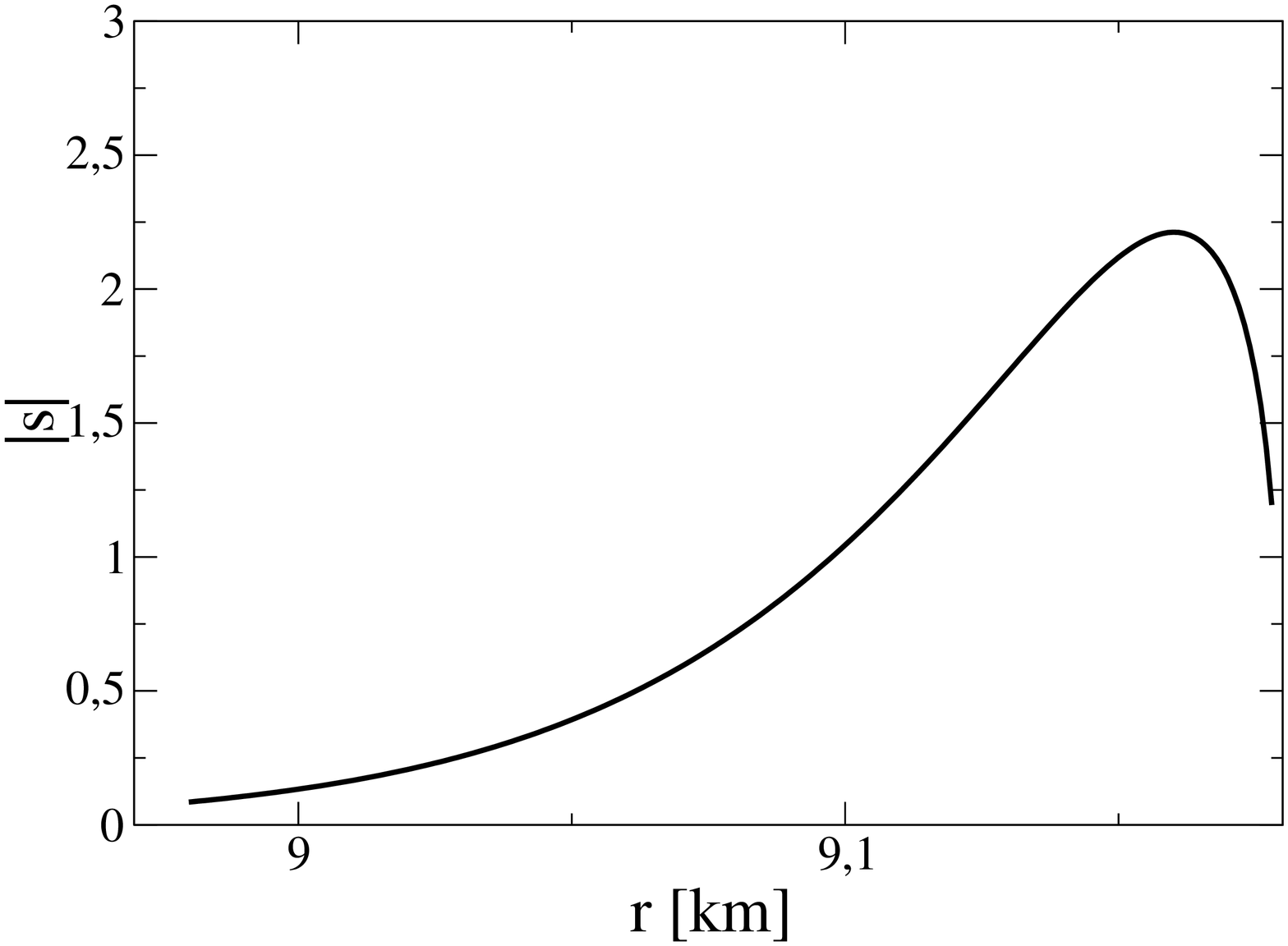} 
\includegraphics[width=8cm]{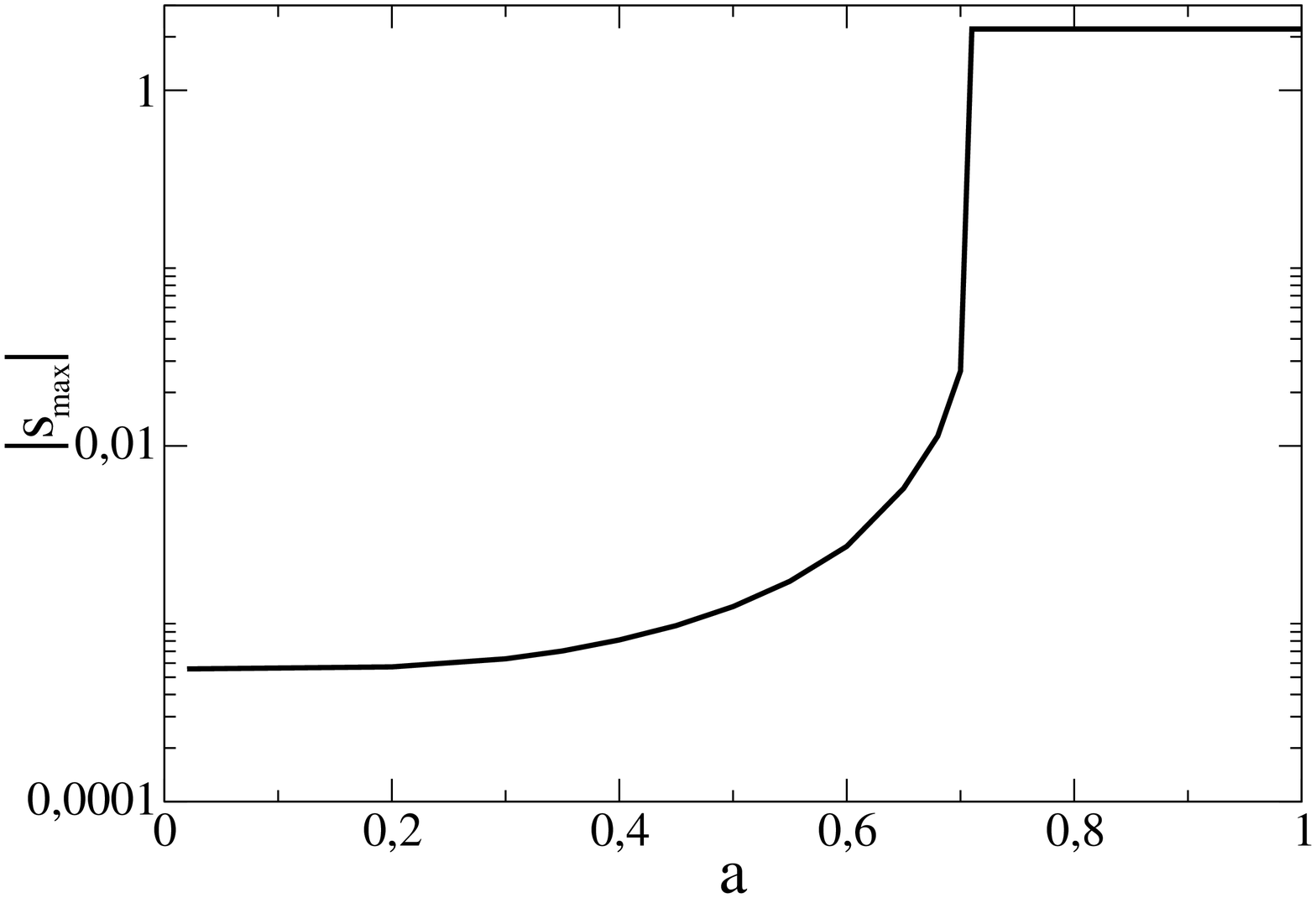} 
\caption{Shear strain produced by the $_1t_1$ oscillation when the inhomogeneous proprieties of the ionic crust are taken into account. Top: Shear strain across the ionic-crust layer  for $a=0.8$. Bottom: Maximum value of the shear strain produced for  different values of the structure parameter $a$. \label{fig:strain}}
\end{figure}

The shear strain corresponding to the ionic crust oscillation is reported in the upper panel of Fig.~\ref{fig:strain} as a function of the radius. Comparing Fig.~\ref{fig:stain_cost}  and the upper panel of Fig.~\ref{fig:strain} one can see that a new feature has appeared. The shear strain does not have a maximum at the CCSC crust - ionic crust boundary. The reason is that the shear modulus of the ionic crust decreases moving toward the surface and it is therefore easily displaced by a torsional oscillation. However, the shear strain at the surface of the star must vanish, because it corresponds to the no-traction boundary condition, thus a maximum is produced close to the star surface.

This characteristic behavior of the shear strain is present also for the CCSC oscillations and the maximum is located where the inhomogeneous term in Eq.~\eqref{eq:Wi} is maximum. In the bottom panel of Fig.~\ref{fig:strain} we show the maximum value obtained for the strain as a function of the parameter $a$.  From this picture is clear that for any value of $a$ the shear strain is large, larger than $10^{-4}$, possibly leading to crust cracking. We will discuss more in detail this issue in Sec.~\ref{sec:conclusions}.

\subsection{Temperature and magnetic field effects}
So far we have neglected the effect of the  temperature and of the compact star magnetic field. In this section we  discuss the range of validity of the presented analysis; in particular we estimate the range of temperature and magnetic field for which the presented analysis is approximately applicable.

\begin{figure}[ht!]
\includegraphics[width=8.cm]{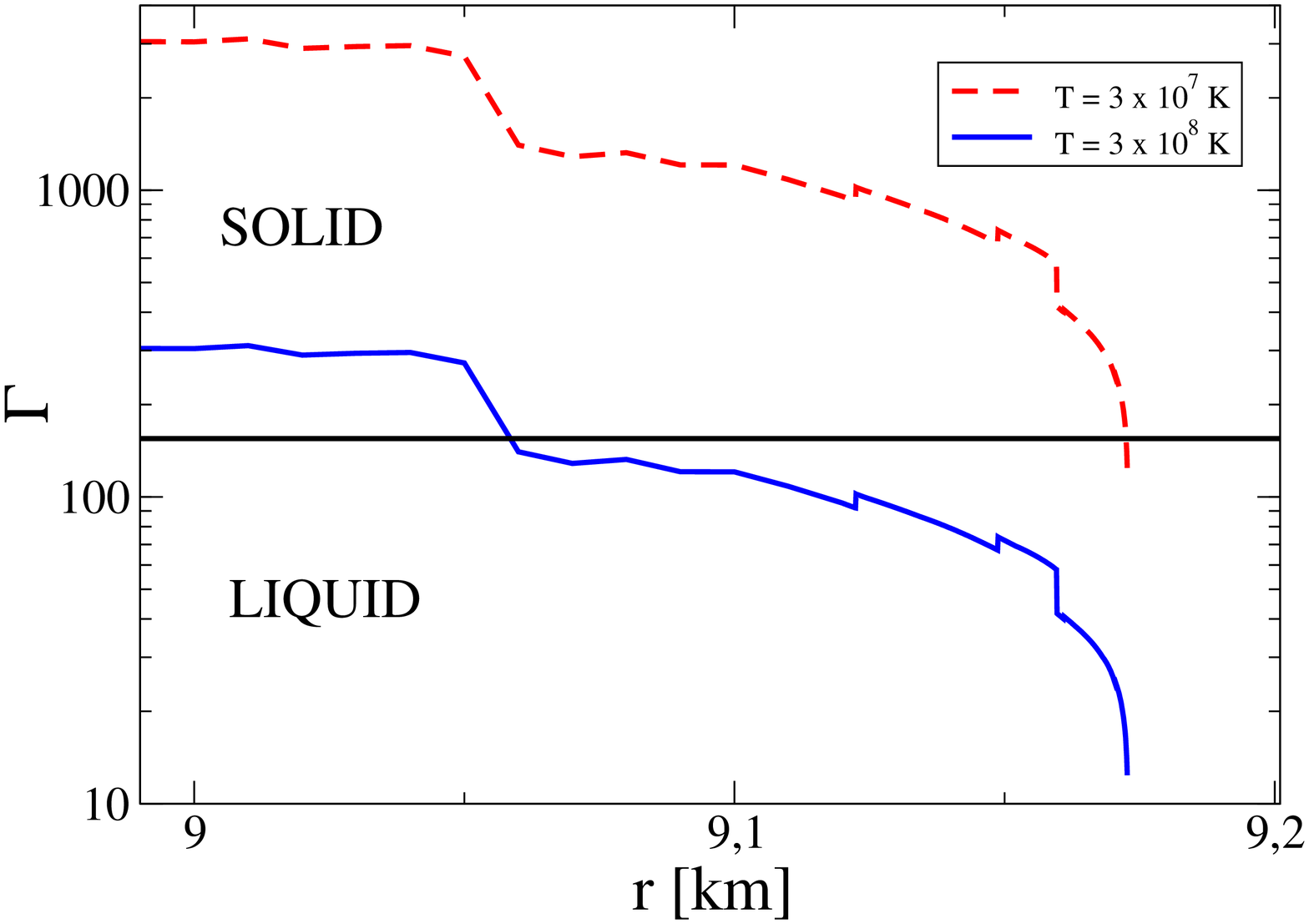} 
\includegraphics[width=8.cm]{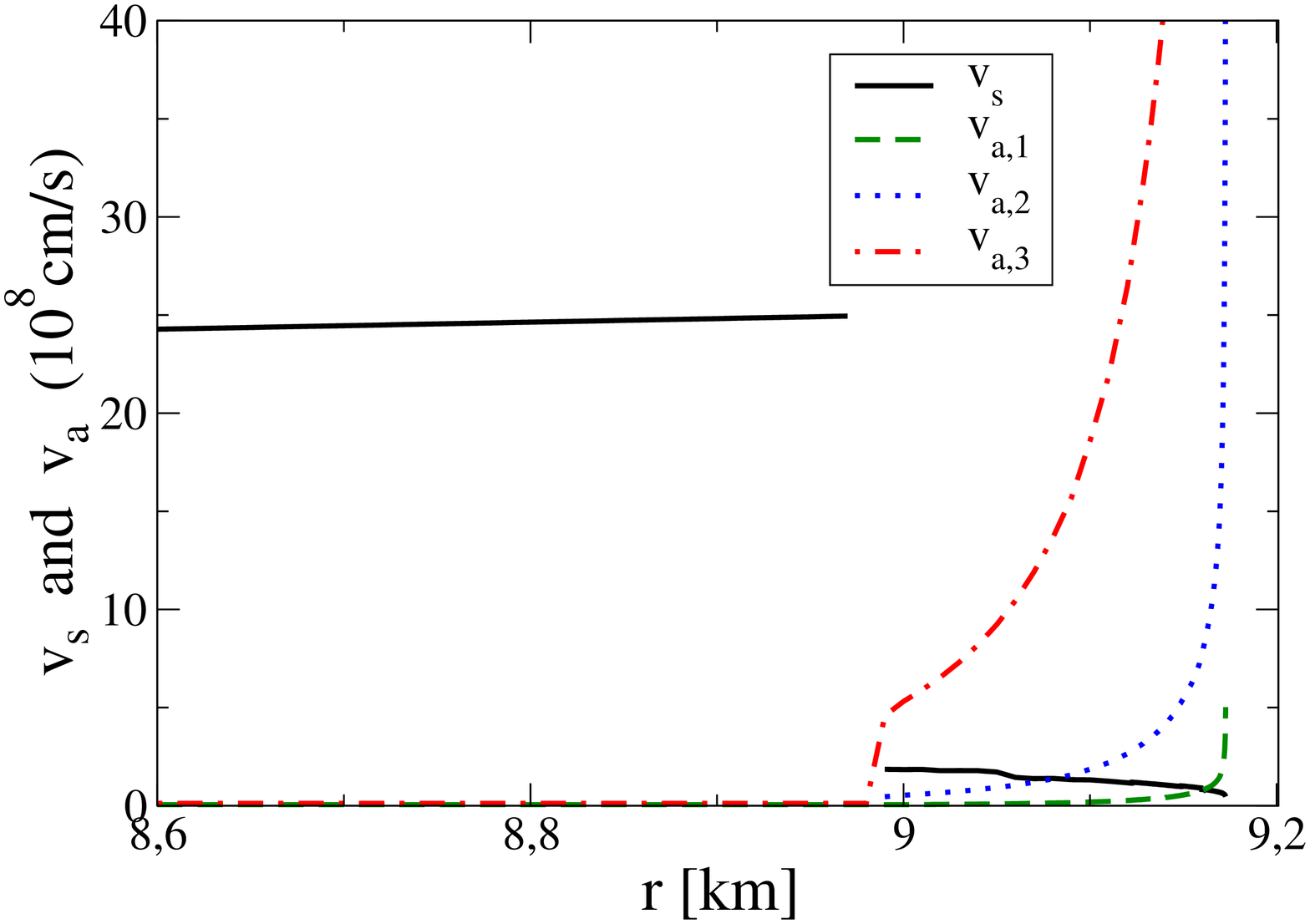} 
\caption{Top: Coulomb parameter for two different values of the temperature. The horizontal line corresponds to $\Gamma=175$, which is  the value that separates the liquid and solid phases.  Bottom: shear and Alfven velocities. The $v_{a,1}$ curve is obtained for $B=10^{13}$ G, the $v_{a,2}$ curve is obtained for $B=10^{14}$ G and the curve $v_{a,3}$ is obtained for $B=10^{15}$ G.  
 \label{fig:gammasansalfven}}
\end{figure}

Regarding the temperature effects, it is known that the shear modulus decreases with increasing temperature,  vanishing at the melting temperature. A typical compact star surface temperature is of about $10^5-10^6$K, but a larger temperature is attained in the interior, see for example~\cite{Shapiro-Teukolsky}. In any case, the typical energy scale of the CCSC crust is of order tens of MeV, thus the temperature effect in the CCSC crust layer can be neglected. On the other hand,  
a fraction of the ionic crust is  believed to be strongly affected by the temperature. We will estimate the temperature at which the present analysis is valid considering the  amount of the ionic crust that is  liquid, forming the compact star ocean. The existence of the ocean is due to the fact that  the  surface temperature of compact stars is larger than the melting temperature of some ionic crust chemical element.   The transition from  solid to  liquid can be semiquantitatively determined by the ratio between the typical Coulomb energy and the thermal energy
\be \Gamma  = \frac{Z(r)^2 e^2}{a(r) T(r)}\,,\ee with the liquid/solid transition  taking place at  $\Gamma = 175 $. Therefore, the radial position of the liquid/solid transition point in the ionic crust depends on the local value of the temperature. Assuming a constant temperature in the ionic crust (which should be a very good thermal conductor), we can easily estimate the radial point corresponding to the solid/liquid phase transition in the ionic crust,  see the upper panel  of Fig.~\ref{fig:gammasansalfven}. For  $T= 3\times 10^7$ K, corresponding to the dashed red line in the upper panel  of Fig.~\ref{fig:gammasansalfven}, only a small fraction of the crust is liquid.  The  value of the star density at which the transition takes place is $\rho\sim 10^7 $g/cm$^{3}$. Therefore, in the present analysis we have assumed that the crust has at  most a  temperature of the order of $10^7$ K. For larger temperatures, indeed, say for  $T= 3\times 10^8$ K corresponding to the solid blue line  in the upper panel  of Fig.~\ref{fig:gammasansalfven}, most of the ionic crust is liquid. 

Regarding the effect of the magnetic field, in the present model it is two-fold. There is a local effect on the net charge distribution located at $r\sim R_\text{q}$ and a standard magnetohydrodynamic  (MHD) effect in bulk.  The local effect corresponds to a deformation of the charge distribution at $R_\text{q}$, and can be properly absorbed in the boundary conditions. Thus, it leads to a shift of the quantized frequencies. Unless one considers magnetic fields  $B \gg 10^{14}$ G  this shift is very small because the electric field energy density in the electron layer is enormous~\citep{Alcock:1986hz}.  The bulk magnetic field   produces a Lorentz  restoring force which competes with the shear stress.  If the Lorentz restoring force is bigger than the shear stress the associated oscillations are Alfven waves. In particular, if the Alfven velocity is much larger than the shear velocity, then   MHD effects are dominant. In the bottom panel of Fig.~\ref{fig:gammasansalfven} we compare the shear velocity with the Alfven velocity for three different values of the magnetic field. For $B\lesssim 10^{13}$G the shear velocity dominates and the presented analysis applies.  Note that Alfven waves are present even if the crust melts.

\section{Conclusions}\label{sec:conclusions}
We have considered a model of nonbare strange star  comprising color superconducting quark matter surmounted by a standard nuclear matter crust. The internal part of the quarksphere  is in the CFL phase; 
the external part of the quarksphere  is in the CCSC phase and is separated from  the ionic crust by a few hundred Fermi thick electron layer. We have determined the background configuration solving the pertinent TOV equation considering a simple parameterization of the EoS of quark matter and a realistic EoS for the description of the ionic crust. We have considered one of the possible stellar structures in the sequence of configurations that solve the TOV equations. Similar results can be obtained considering strange stars with different masses and radii. 

 Both the CCSC and the ionic crust are rigid and we have studied the torsional oscillations supported by  these two electromagnetically coupled crusts. We have classified the torsional oscillations as CCSC crust oscillations and ionic crust oscillations, depending on which part of the structure has the largest displacement. CCSC crust oscillations are completely negligible if the CCSC crust is thin, say less than $\simeq 2$ km, and only ionic crust oscillations are relevant.  We have considered  both a simplified model in which the two crusts are homogeneous and a more realistic model in which the ionic crust density and shear modulus are radial dependent. In any case we have assumed that the CCSC crust is homogeneous, because the density in the quarksphere changes by less than a factor $2$.  We have obtained that the oscillation frequencies of the $\ell=1$ modes are  order of $10$ kHz, and are not very sensitive to the extension of the CCSC crust. The  $\ell=1$  modes correspond to oscillatory twists of the crust and do not conserve angular momentum, therefore these modes are activated by events that  transfer angular momentum to the strange star crust. A typical event of this sort  is  a pulsar glitch  associated to vortex drift from the CFL core to the CCSC crust.  For that reason, we have assumed that the energy conveyed to  the $t$-mode oscillation is of the order of the one of a  Vela-like glitch.  For definiteness  we have assumed that all the energy of a Vela-like glitch triggers one single mode. The obtained deformation of the ionic crust is very large even considering CCSC crust oscillations. If a fraction $\alpha$ of the Vela-like glitch energy is conveyed to the considered $t$-mode, then the amplitude and the shear  deformation must be scaled by $\sqrt{\alpha}$. If $\alpha \sim 1$,  the strain is such that it will  possibly break the ionic crust layer. If the CCSC crust is sufficiently thin to segregate most of the oscillation in the ionic crust, a much smaller energy suffices to produce a large deformation of the  ionic crust: even considering  $\alpha \sim 10^{-3}$ the shear strain on the ionic crust is of order  $0.1$.  Crust cracking happens if the shear strain is larger than the breaking strain, $s_\text{max}$, see~\cite{1976itss.book.....K} for a discussion of the breaking strain in standard materials.   The breaking strain of the ionic crust is highly uncertain, indeed it is not known which is the microscopic mechanism responsible for the nonelastic response. The widely used range of values is  between $s_\text{max} =10^{-4}-10^{-2}$, but values of $10^{-1}$ could be appropriate for perfect crystals without defects~\cite{1976itss.book.....K} as also shown by recent results obtained  by molecular dynamics simulations of Coulomb crystals~\citep{2009PhRvL.102s1102H}. In order to properly asses where and when the crust breaks, it would be important to compute  the breaking strain as a function of the radial distance. The result of our computation is, indeed, that the maximum strain produced by a $_1t_1$ oscillation is located at few tens of meters below the surface of the ionic crust. Therefore, this is the part of the ionic crust that will  likely break during a glitch. 
We are not aware of any observable that might be related to the breaking of the ionic crust at a specific radial distance. Possibly, giant gamma-ray bursts, see~\cite{Thompson:2001ie}, and quasi periodic oscillations, see~\cite{Israel:2005av, Strohmayer:2005ks,Strohmayer:2006py, Glampedakis:2006apa, Watts:2006hk}, might be put in relation with such a phenomenon, but it would be interesting if some specific observable could help to distinguish this breaking process from standard crustal breaks.  

Note that  the maximum of the shear strain  in the present model is much closer to the star surface than in standard neutron stars, see for example~\cite{1988ApJ...325..725M}.  Moreover, in standard neutron stars the energy of the $t$-modes is spread across the entire crust, extending for more than a km below the star surface. By contrast, in the considered model of nonbare strange stars,  all the energy of the $t$-modes is conveyed in a layer few hundred meters thick,  corresponding to the ionic crust at densities below neutron drip. Clearly, the effect of a reduced crust layer is to maximize the deformation.  Considering that in our model we have taken the largest possible extension of the ionic crust layer, we have certainly underestimated the shear strain.

As a final remark, note that if the $t$-oscillations are not triggered by angular momentum transfer to the strange star crust, then $\ell=1$ modes cannot be excited. Then, the low-lying oscillating mode is the $_2t_0$ mode. We find that the oscillation frequency of this mode is of the order of $5$ kHz and it is weakly depends on $a$. Similar results were reported in~\cite{Lin:2013nza}, where a nonbare strange star model with a CCSC core was considered.  Assuming that a Vela-like glitch energy is conveyed to this mode, we find that the corresponding shear strain is suppressed with respect to the $\ell=1$ mode by about three orders of magnitude. We expect that similar results hold for higher values of $\ell$, meaning that assuming that  an equal amount of kinetic  energy is deposited to the modes with different value of $\ell$, the $\ell=1$ mode will produce the larger shear strain. 
\acknowledgments\noindent{\bf Acknowledgments}\\
The research of G.P. is supported in part by SdC/Progetto Speciale Multiasse ``La Societ\`a della Conoscenza in Abruzzo" PO FSE Abruzzo 2007 - 2013.

\appendix
\section{Oscillations of two tightly bound rigid slabs}\label{sec:rigid_slab}
\begin{figure}[t!]
\begin{center}\includegraphics[width=5cm]{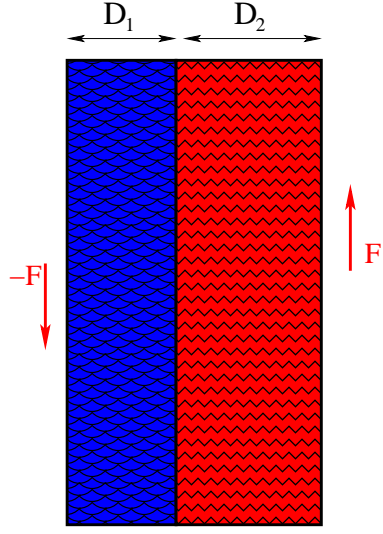} 
\caption{Torque acting of a system of two homogenous slabs in contact by a surface. We assume that the static friction between the two  slabs is so large that the shear oscillations obey the no-slip boundary condition.  \label{fig:twoslabs}}
\end{center}
\end{figure}

Consider the system depicted in Fig.~\ref{fig:twoslabs} consisting of  two slabs with a common contact surface. Suppose that a torque is applied to the structure. In standard conditions the two slabs would easily slide along the common surface with some kinetic friction. If the applied force is below a threshold value, the static friction prevents the relative motion and the whole system  transversally oscillates around the equilibrium configuration. In common materials, the static friction is due to electrostatic Van der Waals forces and is extremely small. Suppose, however, that a strong electrostatic field is present along the contact surface, in such a way that the two surfaces are tightly bound.  Let us consider the fictitious extreme case in which the binding force between the two surfaces is of the same order or large than the force between the atoms in the two slabs. Then, the force necessary to make the two surface slide is larger than the force needed to break the bounds in the bulk of the two slabs. The system of two slabs can be considered as one single nonhomogeneous body with a step-like matter density and shear modulus.   
In these conditions, the transverse displacement field, $u$,  is a continuos function across the surface (no-slip boundary condition)
\be\label{eq:Acontinuity}
u_1(0^-) = u_2(0^+)\,,
\ee  
where we assumed that $z=0$ corresponds to the contact surface. The no-traction boundary condition leads to 
\be\label{eq:Anotorsion}
\nu_1 \partial_z u_1(0^-) = \nu_2 \partial_z u_2(0^+) \,,
\ee
meaning that for unequal shear moduli, $\nu_i$, the derivative of the transverse displacement field is not continuous at the interface (a cuspid point).
Note that the matter densities of the two slabs does not explicitly appear in this equation, indeed the shear moduli  have dimension of a pressure.  The matter density dependence is hidden in $\nu_i$, which in general is a function of the molecular binding force and density. The above no-traction boundary condition basically states that the transverse pressure on the contact surface must vanish.

The frequencies of the quantized oscillations are obtained supplementing Eqs.~\eqref{eq:Acontinuity} and \eqref{eq:Anotorsion} with  no-traction boundary conditions for the free surfaces, which for the considered planar geometry can be expressed as the Neumann boundary conditions $\partial_z u_1(-D_1)=0=\partial_z u_2(D_2)$.  In the linear approximation, a straightforward calculation gives
 \be
(\nu_1  v_{2} + \nu_2  v_{1} ) \sin\left\{\omega\left(\frac{D_1 }{  v_{1}} + \frac{D_2}{  v_{2}}\right)\right\} =(\nu_2  v_{1} - \nu_1  v_{2} ) \sin\left\{\omega\left(\frac{D_2 }{ v_{2}} - \frac{D_1}{ v_{1}}\right)\right\} \,,
\ee
where $v_{i} =\sqrt{\nu_i/\rho_i}$ are the shear velocities. For $\nu_1 v_{2}  \gg \nu_2 v_{1}$ we obtain the approximate solution
\be
2 \sin\left(\frac{\omega D_1 }{  v_{1}} \right) \cos\left(\frac{\omega D_2 }{  v_{2}} \right) =0\,,
\ee
showing that in this case  there are two distinct frequency of oscillations, one determined by the mechanical property of  slab $1$  and one from the property of slab $2$. Note that the frequency quantization 
\be\label{eq:omegan1}
\omega_n = \frac{n \pi v_1}{D_1} \,,
\ee
is the one that is obtained by imposing  Dirichlet boundary conditions for a single slab with width $D_1$ and shear velocity $v_1$. On the other hand,  the frequency quantization 
\be\label{eq:omegan2}
\omega_n = \frac{ (2n -1) \pi v_2}{2 D_2} \,,
\ee
is the one that is obtained by imposing Neumann boundary conditions considering a single slab with width $D_2$ and shear velocity $v_2$. Note that we can recast the condition $\nu_1 v_{2}  \gg \nu_2 v_{1}$ as 
\be
\sqrt{\frac{\nu_1}{\nu_2}} \gg \sqrt{\frac{\rho_2}{\rho_1}}\,,
\ee
and assuming that the slab 1 describes the CCSC crust and that the slab 2 describes the ionic crust, this inequality is certainly satisfied. Therefore, in the planar approximation, the oscillation can be divided in ionic crust oscillation with eigenfrequencies given in Eq.~\eqref{eq:omegan1}, and in CCSC crust oscillation with eigenfrequencies given in Eq.~\eqref{eq:omegan2}. As we have seen in Sec.~\ref{sec:two_homo}, this classification remains approximately valid in the spherically symmetric case.

%\acknowledgments
\bibliographystyle{apj}
%\bibliographystyle{h-physrev4}
%\bibliography{CSC}

\end{document}